# TwiInsight: Discovering Topics and Sentiments from Social Media Datasets

Zhengkui Wang[1], Guangdong Bai[1], Soumyadeb Chowdhury[1]

Quanqing Xu[2] and Zhi Lin Seow[3]

[1] InfoComm Technology, Singapore Institute of Technology, Singapore
*{Zhengkui.Wang, Guangdong.Bai,* Soum.Chowdhury}@singaporetech.edu.sg

[2] Data Storage Institute, A*STAR, Singapore
Xu_Quanqing@dsi.a-star.edu.sg

[3] School of Computer Science, University of Glasgow, Scotland
*Zhilin.Seow@student.gla.ac.uk*

**Abstract.** Social media platforms contain a great wealth of information which provides opportunities for us to explore hidden patterns or unknown correlations, and understand people's satisfaction with what they are discussing. As one showcase, in this paper, we present a system, TwiInsight which explores the insight of Twitter data. Different from other Twitter analysis systems, TwiInsight automatically extracts the popular topics under different categories (e.g., healthcare, food, technology, sports and transport) discussed in Twitter via topic modeling and also identifies the correlated topics across different categories. Additionally, it also discovers the people's opinions on the tweets and topics via the sentiment analysis. The system also employs an intuitive and informative visualization to show the uncovered insight. Furthermore, we also develop and compare six most popular algorithms – three for sentiment analysis and three for topic modeling.

## 1 Introduction

Today, there are a huge variety of social media platforms (i.e., Twitter[1], Facebook[2] and Instagram[3]) which allow users to connect with friends and share various kinds of information like personal daily event, their opinions or complains on a variety of topics and discuss current global issues. These social media data contain a great wealth of information and provide us the opportunities to explore hidden patterns or unknown correlations, and understand people's satisfaction with what they are discussing for different various topics. While many of these social media data is publicly available, it remains challenging to analyze the data to mine the useful information.

---

[1] https://twitter.com
[2] https://facebook.com/
[3] https://www.instagram.com

*Consider one scenario.* A user wants to understand that within different categories (i.e., food, healthcare, transport, technology and sport), what are the most popular topics that people are discussing online? These topics under different categories can help the user better understand the main concerns of people's living perspectives. Additionally, he/she is also interested in identifying the correlated topics across different categories. This helps understand people's interests pattern, like what are the topics people talk about in food category when they talk about one topic in healthcare. Furthermore, the user is also interested in knowing people's satisfaction or opinion when they talk about one topic. Given this scenario, the social media data becomes an important resource that can be analyzed to provide the solution.

There are many existing research works working on detecting the topics via the topic modeling from the social media data [11][12][13]. These research aims to discover the underlying key topics that occur in a set of online posts which assists companies to monitor and summarise information that people are discussing on the social media platforms. Another group of research have been focusing on identifying the people opinion via the sentiment analysis from the social media data[6][7][9][10]. These research aims to identify users sentiments (e.g., Positive, Negative or Neutral) based on what they have shared online to understand people's satisfaction on each topic. However, most of these works do not classify the data according to their categories, but analyze the data as one whole.

In this paper, taken Twitter data as one example, we present **TwiInsight**, one system developed to identify the insight from twitter data. TwiInsight aims to automatically detect the topics and their sentiments under different categories, and further to identify the correlated topics and trends across-categories. TwiInsight facilitates one generic framework including the data collection, data analysis and data visualization modules. In particular, the data collection module employs various data crawlers to retrieve online tweets which are further stored in MongoDB[4], one NoSQL database in a real-time manner. To distinguish the tweets from different categories, the crawler crawls the tweets by using the popular hashtags related to each category. Based on the tweets crawled within each category, TwiInsight enables three different types of data analysis (namely topic detection, sentiment analysis and correlation detection) and one user-friendly visualization.

The contribution of the paper is two-fold: First, we present a system design that is able to integrate different technological opportunities (information extraction tools) to showcase how the social media data can be utilized for decision making. Second, we develop and compare different algorithms (three algorithms for sentiment analysis based on standford CoreNLP, Haven On Demand(HOD) and Monkeylearn; three algorithms for topic extraction based on Skeyttle, Rapid Automatic Keyword Extraction (RAKE) and GATE Twitter Part-of-Speech Tagger; and another algorithm for correlated topics detection based on co-occurrence matrix and TF-IDF algorithm) to validate and evaluate our design decisions and choices.

---

[4] https://www.mongodb.com/

The rest of the paper is organized as follows: We will provide the related work in next section, and introduce the system and algorithms design in section 3. Section 4 presents the experimental evaluation and section 5 concludes the paper.

## 2    Related Work

This work is inspired and related to multiple groups of research. In this section, we summarize and briefly discuss them.

**Twitter Data Analysis Tools.** There are a few tools available for analysing social networking data for different application scenarios.

*Keyhole*[5] offers an extensive number of packaged analytics visualisations that illustrate metrics in an easy-to-read graphs and layouts for keywords, account summary and so on. It provides a variety of dashboards to indicate the results according to user input hashtag as the search key.

*Tweet Sentiment visualization*[6] is an analytics application developed to study ways to visualize sentiment for unstructured and also non-grammatical tweet. It offers a comprehensive suite of sentiment visualization techniques that use searched keywords to analyze the sentiment behind each tweet associated with the searched keywords.

*Twitter Analytics*[7] is another analytics application that is developed by Twitter. It has two main tools. One is Tweet activity dashboard which allows user to learn more about their Tweets and understand their audience. The other is an enhanced analytics known as audience insights, which provides a more de-tailed breakdown of user's followers to help advertisers better strategise their advertisement.

**Data Crawlers.** There are a few crawling systems which have been used in the past few years to support Twitter research. Song et.al [1] explored topological and geographical properties using Twitter. Using REST API methods, they extracted tweets from April 1st to May 30th 2007 and obtained around 1.3million tweets from 76k users. For the period that the authors crawled, Twitter had just started up such that it is insufficient to collect a significant amount of data. Several other researchers crawled Tweet from Twitter to investigate sentiment analysis [2], to develop spam detection system to identify suspicious users [3] and to detect critical events promptly [4]. Most of the research were systems that is focusing on specific data.

**Sentiment Analysis.** Sentiment analysis is a growing area of Natural Language Processing. It can be handled at many levels of granularity, starting from being a document-level classification [5] to sentence-level classification [6] and more recent at phrase-level classification [7].

Given that Tweet has 140-characters limit, it is the most similar to sentence-level sentiment analysis when analysing the sentiment result of the tweet [8]. However, tweets are informal and highly unstructured as well as the nature of social media

---

makes Twitter sentiment analysis a different task. It is an open question on the accuracy of sentiment analysis on well-formed data as compared tweets that are highly unstructured and also non-grammatical [9].

For social media data, it is common that people use emoticons or acronyms or both to express their emotions. With this distinct feature of social media data, many researchers using this as one of the factors to improve the accuracy of the sentiment analysis result. Some of the early results that include an emoticon into their sentiment analysis are produced by [9] and [10]. The former [9] builds models using Multinomial Naive Bayes (MNB) and Support Vector Machines (SVM). Instead of just using emoticon to classify tweet, [10] first created emoticon and acronym dictionary to preprocessed the data. Using processed data, it classifies tweet based on the result of prior polarity of words.

**Topic Modeling**. Topic modeling is a critical topic in text mining. The most commonly used tool for topic modeling is Latent Dirichlet Allocation (LDA) [11], which has been extended for social networking data analysis. Hong et.al [12] propose several schemes to train LDA and compare their quality and effectiveness. Zhang et al. [13] propose to incorporate LDA into community detection. McCallum et al. [14] proposed a model to discover groups among the entities and topics

**Correlated Topic Detection**. Correlated Topic Detection is to detect patterns or trends on a set of topics that occur frequently together in a tweet over time. [15] models the correlation between news and tweets. [16] uses surveys on political opinion and tweets that was posted in the same period of time to analyse the correlation between sentiment keywords and the result of the poll. [17] makes use of the sentiment of each topic to compute the Pearson correlation.

## 3    TwiInsight System and Algorithm Design

In this section, we will first introduce the problem studied and system architecture overview followed by the data collection introduction. We will further present all the developed algorithms for sentiment analysis, topic extraction and correlated topics detection, followed by the visualization discussion.

### 3.1    Problem and System Architecture Overview

The problem studied in this paper is to understand what people are discussing and what is people's view/attitude over different categories (e.g., food, healthcare, transport, sports and technology), and to detect patterns or trends on topics that occur frequently together across these categories.

To address this problem, we propose one automatic solution which is achieved by analyzing the social media data. To showcase the idea, we introduce TwiInsight, one system developed to discover the insight of the Twitter data. Though TwiInsight is developed over the twitter data, it can be easily extended to analyze different social media data sources.

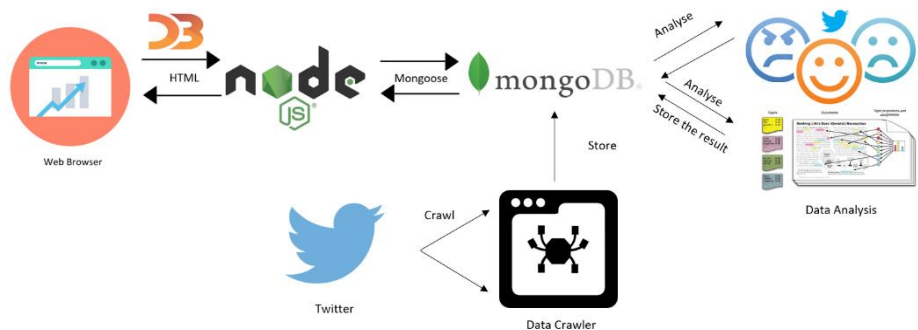

Figure 1: TwiInsight System Architecture

Figure 1 provides the system architecture overview of TwiInsight. The system consists of three main components including data collection, data analysis and data visualization. The data collection module is developed to crawl the tweets from Twitter using data crawlers and to store the tweets into MongoDB, a NoSQL database for scalability and scheme less data storage purpose. After getting the tweets, the system mainly performs three different types of analyses to answer the following questions:

- What are people's opinion on the specific topics to help us understand their satisfaction of those topics?
- What are the topics discussed by people online to help us understand people's interests?
- What are the correlated topics that occur together online to help us understand the trends and patterns of user's interest?

Another component of TwiInsight is one data visualization module which displays all the findings discovered to users. This enables decision makers to easily digest the information for a better decision making.

### 3.2 Data Collection

To collect the tweets data, data crawlers are used in the system to automatically crawl the updated tweets including the tweet text, time, location and so on every day. To showcase the idea, we focus on English tweet and choose five categories which are Food, Healthcare, Sport, Technology and Transport. Note that the system can be easily extended to apply different crawlers to crawl the tweets in different languages and categories. The data crawlers are developed by using the Twitter4J, a Java library for the Twitter Application Programming Interface (API) that can easily integrate data crawler with Twitter service.

For each category, we gathered 50 or more popular hashtags which are used to crawl the related tweets. A hashtag is a convention among Twitter users to create and follow a thread of discussion by prefixing a word with a '#' character. Table 1 provides some sample hashtags used in the system.

**Table 1.** Hashtags Used in Each Category.

| Category | Hashtags |
|---|---|
| Food | snack, cleanrecipes, dinning, cleanRecipe, eatclean, organic, protein glutenfree, vegan, fitfood, eatclean, resturant, café, calories, cook |
| Healthcare | PlasticSurgery, digitalhealth, MedicalHumanities, pharma, rehab, pharmacy, SLP2B, chiropractic, Migraine, BCSM, Diabetes |
| Sport | sport, WorldCup, football, soccer, basketball, exercise, ball, yoga, workout, training, treadmill, gainz, workout, getfit, justdoit, geekabs |
| Technology | technology, tech, technews, techno, IoT, innovation, BigData, ArtificialIntelligence, AI, Digital, VirtualReality, CloudComputing, IT |
| Transport | Transportation, automobile, sustainable, traffic, TrafficJam, selfdriving, civilengineer, uber, smrt, sbstransit, traffic, LTA, transport |

When the data crawler crawls tweets from Twitter, it needs a reliable platform to store and retrieve large amount of data easily. Visualizing big data also needs a database that can process large volumes of data without dropping the performance time to load the graph. Additionally, every tweet tweeted by Twitter user varies in terms of hashtag, language and topics. Based on these requirements, MongoDB is chosen as the backend in this system, as it has flexible schema in storing data. The collections do not enforce each document to have the same structure with one another.

### 3.3 Topic Extraction

To understand the topics that people are discussing on each tweet, TwiInsight employs the topic modeling techniques. Topic modelling is also sometimes known as keyword extraction, term extraction or keyword analysis. It is done by finding a single word or phrase that describes the main notions and entities mentioned in the text.

To enrich the study, we develop and evaluate three different algorithms to extract the topics of each tweet. These algorithms are designed based on different models: (1) Skyttle[8] (2) Rapid Automatic Keyword Extraction (RAKE) [18] and (3) GATE Twitter Part-Of-Speech Tagger[9]. To improve the topic modelling performance, we first clean the data by removing all the stopwords, username and URL and so on.

**Skyttle-based algorithm**. The first topic extraction program we developed is based on the Skyttle which provides text analytics services to extract interesting information from text and returns the result in a structured format for in-depth data analysis [19]. Skyttle runs on the Mashape's infrastructure[10], which is the largest API marketplace and management network. It is working as one Software as a service (SaaS) system where the developer can access the API using a Mashape key. To use Skyttle, we let the program read the tweets from MongoDB one by one and return the keywords from each tweet for further usage. For illustration, let us take one tweet "Tweet Example 1" as one example.

---

[8] http://www.skyttle.com/

[9] https://gate.ac.uk/wiki/twitter-postagger.html

[10] https://www.mashape.com

*Tweet Example 1:* "You are my new fav chef Kevin Belton. Made shrimp boil and it killed. # cook # food".

Based on the Skyttle algorithm, Figure 2 provides one sample result of processing tweet example 1, where terms are those identified keywords and count indicates their frequency in the text.

```
{
  "docs": [
    {
      "terms": [
        {
          "count": 1,
          "term": "chef",
          "id": "af74d302b526b714ceb624cacd0bb9670b0462d0"
        },
        {
          "count": 1,
          "term": "Kevin Belton",
          "id": "84f4966b50704820129d35025588feac9a9c1485"
        },
        {
          "count": 1,
          "term": "shrimp boil",
          "id": "af74d302b526b714ceb624cacd0bb9670b0462d0"
        },
        {
          "count": 1,
          "term": "food",
          "id": "af74d302b526b714ceb543cacd0bb23521b0462d0"
        }
      ],
      "language": "en"
    }
  ],
  "warnings": []
}
```

Figure 2: Sample output of Skyttle algorithm

**Rapid Automatic Keyword Extraction (RAKE)-based Algorithm.** The second topic extraction algorithm we developed is based on RAKE. RAKE is a well-known algorithm implemented in Python for extracting keywords from text[18]. It is an algorithm that is category-independent and language-independent for extracting keywords from text. The algorithm works by extracting all the non-stopwords and then scoring these phrases across the text. Unlike other algorithms, it does not remove punctuation signs and instead treated as sentence boundaries. It also uses one stopwords list where the stopwords are treated as phrase boundaries to help generate keywords or phrases that consist of one or more non-stopwords. After that, the algorithm computes the properties of each extracted keyword or phrase which is to sum the scores for each of its words. It is scored according to their frequency and the length of the phrase in which they appear. We develop one Java version of RAKE in the system. Figure 3 indicates the result of processing the tweet example 1 based on the RAKE algorithm.

```
Keyword: #cook #food, Score: 4.0
Keyword: made shrimp boil, Score: 9.0
Keyword: fav chef kevin belton, Score: 16.0
Keyword: killed, Score: 1.0
```
Figure 3. Sample output of RAKE algorithm.

**GATE Twitter Part-Of-Speech Tagger-based Algorithm.** The third program is developed based on the Gate Twitter Part-of-Speech (POS) Tagger which is one state-of-the-art tagger for tweets data. The tagger aims to achieve competitive accuracy and was developed using the Penn Treebank tag set so that it can integrate the tagger into any tools seamlessly. The tagger is an adapted and augmented version of a leading

Conditional Random Field (CRF) -based tagger, customised for English tweets [20]. It stated that the tagger achieved 91% accuracy on tokens on their evaluation set which is considered very high for tweets. Most importantly, it has a relatively high accuracy on whole-sentence correct. For tasks like dependency parsing and event extraction, it is crucial to achieve good performance for getting the whole sentence.

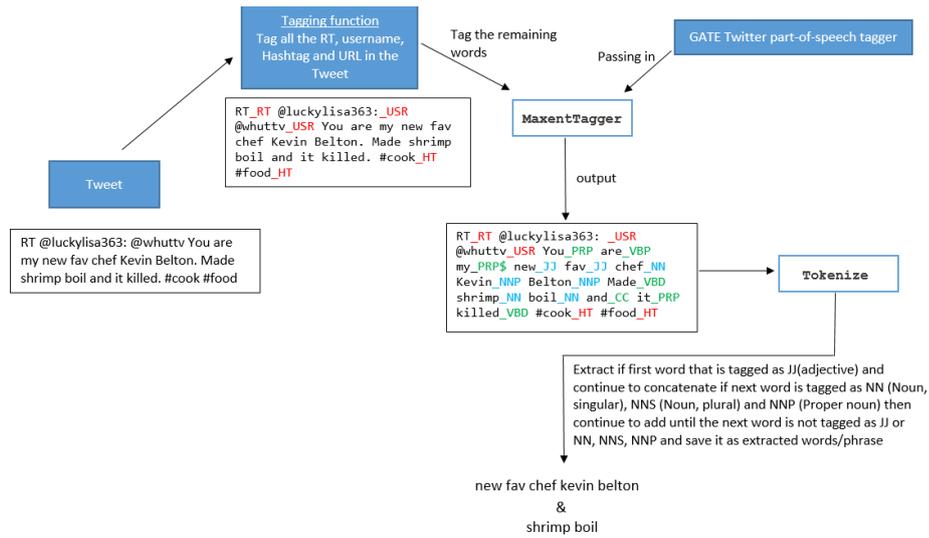

Figure 4. Procedure of Integrating GATE Twitter POS Tagger for topic modeling

Figure 4 illustrates how to integrate GATE Twitter POS tagger with topic modelling to obtain a list of topics or keywords. Our developed topic modeling algorithm works as follows: it takes in tweets and passes it to a tagging function to tag all the RT, username, hashtag and URL. With the output, it will pass to maxentTagger with GATE Twitter POS tagger model which will perform part-of-speech for the rest of the words. When the GATE Twitter POS tagger produce the output, it will then undergo tokenization to select words that are tagged with JJ or NN (Noun, singular), NNS (Noun, plural) and NNP (Proper noun) and continue to add the position of the word. If the next word is tagged as NN, NNS, NNP it will then continue to do so until the next word is not tagged as JJ or NN, NNS, NNP and save it as extracted words/phrases.

### 3.4 People Opinion Analysis

To understand people's opinion or satisfaction on any specific topics, TwiInsight employs the sentiment analysis technology which enables automated process of understanding whether the text is positive, neutral, or negative. Three algorithms are developed in this paper to extract the sentiments of each tweet based on different models: (1) StanfordCoreNLP [21] (2) Haven OnDemand[11] and (3) Monkeylearn[12].

All the stopwords, username and URLs are all removed, before processing the tweet to all three approaches.

**Stanford CoreNLP-based Algorithm.** The first program we developed to analyze the tweet sentiments is based on StanfordCoreNLP. StanfordCoreNLP is a Java library for natural language analysis tools. Stanford CoreNLP can integrate many NLP tools, including the part-of-speech (POS) tagger, the parser, sentiment analysis and many other more tools [21]. Sentiment analysis is usually performed by defining a sentiment dictionary, removing stopwords, tokenizing the tweet, scoring for individual tokens and aggregating the scores to finalise a sentiment score. Instead of performing several different numbers of tasks, StanfordCoreNLP's sentiment model first identify the sentiments uses phrases to instead of words. It then builds a sentiment tree to compute the overall sentiment. The sentiment model is pre-trained on approximately 12,000 sentences on a Recursive Neural Tensor Network [22].

The algorithm works by using a CoreNLP pipeline to define annotators to build annotations over a stream of text. Annotations is to define how the tweet to be analysed or annotated (i.e., how the tweet should be processed by NLP tasks). For sentiment analysis, annotators are defined to tokenize the text (tokenize), splits a sequence of tokens into sentences (split), performed basic syntactic analysis (parse) and sentiment polarity detection (sentiment).

After initialising the pipeline with the annotators, tweet will then be passed into the pipeline for processing. The processing is to build a sentiment tree to compute the overall sentiment. Figure 5 shows how the sentiment tree will look like. After computing, the sentiment analysis will return a sentiment score instead of the sentiment label e.g. Positive. Therefore, an array was created to convert the sentiment score to sentiment label.

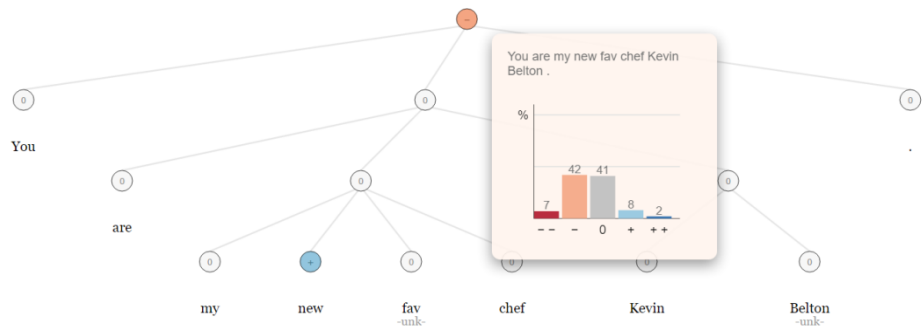

Figure 5. Sentiment Tree: Each leaf node indicates the sentiment value for each word in the sentence.

Haven On Demand (HOD) is a platform that supports an array of data processing API like sentiment analysis, image analysis, predictive analysis and many others. It offers great flexibility to integrate with native applications.

The sentiment analysis model works by recognising the nouns, verbs, and other parts of speech through matching the structure of the tweet. From then, the tweet is classified based on the classification of the constituent words as positive, negative, or neutral. Classification is done by comparing the words in the tweet with a dictionary of positive and negative words of different types with a score and then return a calculated sentiment score. Based on the sentiment score, it will assign sentiment polarity accordingly. It can intelligently identify both the topic and the sentiment.

**Haven OnDemand-based Algorithm.** Another sentiment analysis program we have developed is based on the Haven OnDemand model. The Haven OnDemand Sentiment Analysis API uses URL to take in Tweet input that can be provided in the text parameter [23]. It allows developers to provide multiple inputs by specifying multiple input parameters. Based on the tweet input, the algorithm extracts positive and negative phrases. Each of the sentiments extracted contains some valuable information that determines the sentiment. The sentiment is measured by a score value which indicates the strength of the sentiment. If there is more than one positive or negative phrase identified, the algorithm will then perform an aggregation, which then takes into account of all the sentiments extracted and calculate the final sentiment result and score for the Tweet input.

**Monkeylearn-based Algorithm.** The third sentiment analysis program developed is based on the Monkeylearn's English Tweets module. MonkeyLearn is a platform that used machine learning to get relevant data from text. MonkeyLearn aims to retrieve and classify information from text for specific needs, and integrate it into their own platforms and applications in an easy, fast and cost-effective way. Like Stanford coreNLP, MonkeyLearn also provides the option for user to create their own customised classifier to train their model using MonkeyLearn's machine learning algorithms in MonkeyLearn's cloud. The developed algorithm uses a customize classifier that classifies tweets in English according to their sentiment polarity. The classifier was trained with approximately 21,000 tweet samples in MonkeyLearn's cloud.

### 3.5 Topic Correlation Detection

To detect patterns or trends on the topics that occur frequently together over time, TwiInsight employs one topic correlation detection Java program that runs through all the topics and build a matrix based on their frequency occurrence. The co-occurrence matrix captures topic relationships and hence harnesses the semantic information on topics co-occurring in the tweets collection. Co-occurrence information can be used to recommend topics or get the top occurrence topics.

By nature, popular topic will co-occur highly with all topics. In order to resolve this problem, computing the discriminating power of a given topic is important in order to give a "weight" the topics. For example, if topic A co-occurs 45 times with a rare topic B (which was only discussed in 100 tweets) and 120 times with a popular topic C (which was discussed in 12,000 tweets) topic B should be more relevant correlated topic to topic A rather than topic C. To "weight" the topics, a weighting function was introduced to discount popular topics and ranked those topics with highest co-occurrence without thinking that it is highly correlated or just because it correlated to the popular topic. To achieve this, we adopt the well-known TF-IDF[13] (Term Fre-quency-Inverse document frequency) as the weighting factor. Due to the space limitation, the detail algorithm information is omitted here.

---

[13] http://www.tfidf.com/

## 3.6 Data Visualization

To enable an intuitive information exploration, TwiInsight employs multiple data visualizations for users. After data analysis, the system would show all the finding via one web GUI. Here, we briefly present the main visualizations that the system supports.

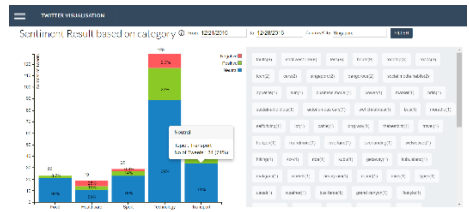

Figure 6. Aggregated sentiments with topics

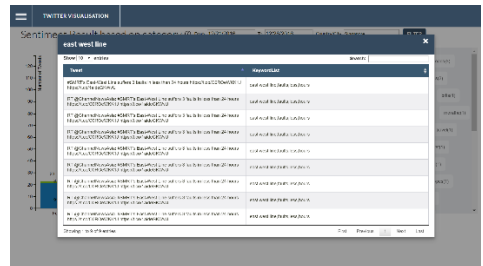

Figure 7. Tweets related to each topic

Figure 6 indicates the stacked bar with all the aggregate sentiment values for each category, where the blue, green and red parts are the percentage of positive, neutral and negative tweets. Meanwhile, the user can filter the data based on different dimensions such as time and location. When the user mouse over the stacked bar, it shows detail sentiments value. The right panel can dynamically show all the topics that extracted under the particular topic once the user clicks the bar in the left. The system also allows users to view all the tweets that related to one topic by simply clicking the topics as shown in Figure 7.

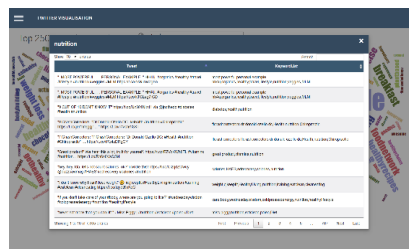

Figure 8. Extracted topics and its tweets

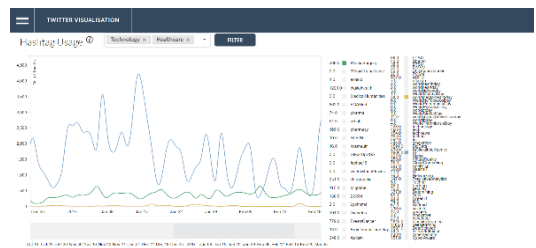

Figure 9. Hashtag usage patterns over time

To assist users to explore the extracted topics, TwiInsight uses the word cloud to list all the topics under each category as shown the Figure 8. When users click one particular topic, all the related tweets will be shown in another popup window. The system also enables users to see the trends of the hashtags over time by using the filtering as shown in Figure 9.

Additionally, TwiInsight has also provided a line graph to indicate the top *n* most correlated topics under different categories over time as shown in the Figure 10. This helps understand the trends and patterns of how to correlated topics evolves. Furthermore, to view the location difference, it also provides an information summarizer based on the google map as shown in Figure 11.

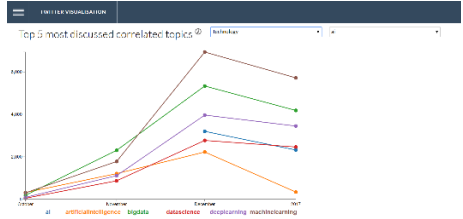 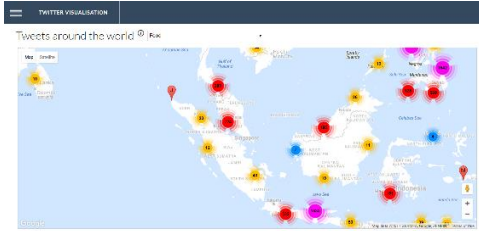

Figure 10. Correlated topics over time        Figure 11. Information summarization based on geo-location

## 4        Performance Evaluation

In this section, we present performance evaluation for the sentiment analysis and topic extraction algorithms based on real dataset collected from Twitter. All experiments are conducted on a Windows server with an Intel(R) Core(TM) m3-6Y30 1.51GHz processor and 8GB of RAM, running 64-bit Windows 10 Home.

### 4.1        Sentiment Analysis Algorithms Comparison

There are many discussions to be found on how well the sentiment analysis performed and the various techniques that measure it. Sentiment analysis is considered as a subjective task, it is not only difficult for an algorithm to compute but also for humans. If different people are tagging for 1,000 tweets its respective sentiment polarity (positive, neutral, negative), most likely they do not agree the sentiment result for most of those tweets. It may vary a lot for human agreement on the sentiment result of on a tweet because of the subjective nature of sentiment analysis. We believe that there is no sentiment analysis that achieves 100% accuracy, but it is still useful when trying to get a macro-level feel for the sentiment of a topic or set of topics from tweets.

**Table 2.** Some of the Sentiment Result of all Three Algorithms

| Tweets | Stanford based | Haven OnDemand based | Monkey-learn based | Evaluated Sentiment based |
|---|---|---|---|---|
| RT @Alex Verbeek: Australian seaweed found to eliminate more than 99% of cow burp methane https://t.co/VfjWZ0KRJp #climate #food | Negative | Neutral | Positive | Positive |

| #RT [Homemade] The biggest taco I have ever done #food https://t.co/doibNlnQsC | Positive | Neutral | Positive | Positive |
|---|---|---|---|---|
| #RT [Homemade] Chicken Curry Hand Pies! #food https://t.co/uUyRAcBB5t | Positive | Neutral | Neutral | Neutral |
| RT @veganposters: Most importantly, don't feel guilty if you decide to have an unhealthy treat - treats... – Whitney Lauritsen #vegan https | Negative | Negative | Negative | Positive |
| RT @PengPengPeny: Can't live without cheese? They can't live with it. All milk is mothers milk. #DigOutYourSoul GO #vegan End cruelty. http | Neutral | Neutral | Negative | Negative |

Table 2 shows the sample sentiment results of each algorithm and the sentiment result of the evaluation. Note that due to space limit, we do not show the full comparison of all the sentiment results. To calculate the accuracy of each algorithm, we count the number of tweets in each algorithm that was tagged the same as the evaluated sentiment result, where the evaluated sentiment results are correct decisions manually given to the tweets via human input. The percentage of accuracy for each algorithm was calculated using the number of tweets tagged the same as the evaluated sentiment result over the total tweets. As shown on Figure 12, the accuracy of MonkeyLearn-based algorithm is 60%, which is much higher than the rest: Haven ondemand-based (35%) and Stanford CoreNLP-based (40%). With that, MonkeyLearn-based algorithm was finally deployed for sentiment analysis in our system prototype.

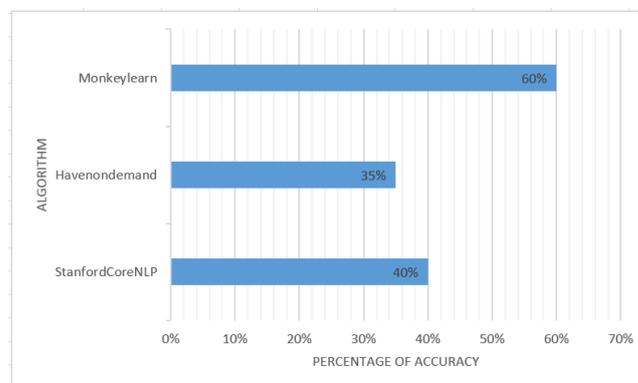

**Figure 12.** Accuracy Percentage of all the Algorithms

### 4.2 Topic Modelling

Like the evaluation for sentiment analysis, twenty participants took part in this evaluation. They were given a tweet and three sets of topics extracted by the three algorithms. They need to pick the best topics among the three sets of topics and based on that to determine on average which algorithm produces the best result. Table 3 shows the number of participants chose that algorithm's topics as best for each tweet as some example. Note that due to space limit, we do not present the full comparison of all the Topic Modelling results.

**Table 3.** Some of the Topic Modelling Results of all Three Algorithms

| Tweet | Skyttle-based Alg. | Rake-based Alg. | Gate Tagger-Based Alg. |
|---|---|---|---|
| @JoeBugBuster A6 #retro #food of course! #nostalgiachat | 14 | 2 | 4 |
| RT @PengPengPeny: Can't live without cheese? They can't live with it. All milk is mothers milk, #DigOutYour-Soul GO #vegan End cruelty | 0 | 0 | 20 |
| #RT [Homemade] The biggest taco I have ever done #food https://t.co/doibNlnQsC | 3 | 1 | 16 |
| #RT [Homemade] Chicken Curry Hand Pies! #food https://t.co/uUyRAcBB5t | 4 | 0 | 16 |
| #RT [Homemade] Girlfriend made her famous pumpkin pie today #food https://t.co/F5HZnLfLyz | 16 | 0 | 4 |

For each tweet that has the highest number of participants, it means that the algorithm produces the best result. From the evaluation, we have identified that for 80% of all the twenty tested tweets, GATE Twitter POS tagger-based algorithm was ranked as the best algorithm. And for 20% of the tweets, Skyttle-based algorithm was ranked as the best. And RAKE-based algorithm was not ranked the best in any of the tweets. This indicates the GATE Twitter POS tagger-based algorithm performs the best in terms of the topic extraction. This is reasonable as it utilizes one Twitter customized tagger. The RAKE-based algorithm performs the worst which may be because it does not identify the punctuation as a part of the actual phrase that decreases the accuracy.

## 5    Conclusion

While social media platforms contain a great wealth of information, it remains challenging to analyze them for different application purpose. In this paper, as one showcase, we presented TwiInsight, one system integrated different technological opportunities for insight discovering over Twitter data. TwiInsight facilitates one data collection module to crawl online tweets under different categories in a real-time manner. It further analyzes the tweets to automatically detect the topics, sentiments and correlated topics over time and employs different visualizations after information integration. In addition, we evaluated and compared multiple algorithms for each analysis component for different design decisions and choices.